\documentclass{aa}
\usepackage{times}
\usepackage{graphics}
\usepackage{amsfonts}
\begin{document}

   \thesaurus{11
              (11.19.2;
               11.19.6;
               05.03.1)}
   \title{A new approach to the problem of modes in the Mestel disk}

   \author{M.~Demleitner
          \and
          B.~Fuchs
          }

   \offprints{M. Demleitner}

   \institute{Astronomisches Recheninstitut, M\"onchhofstr.~12-14, 69120
            Heidelberg
          email: msdemlei@ari.uni-heidelberg.de}

   \date{Received ????; accepted ????}

   \maketitle

   \begin{abstract}
We examine the modes admitted by the Mestel
disk, a disk with a globally flat rotation curve.  In contrast to
previous analyses of this problem by Zang (\cite{1976PhDT........26Z}) and 
Evans~\&~Read (\cite{1998MNRAS.300...83E}, \cite{1998MNRAS.300..106E}), we
approximate the orbits to obtain almost closed expressions for the
kernel of the integral equation governing the behaviour of the modes.
Otherwise we, like them, follow Kalnajs' programme 
to simultaneously solve the Boltzmann and Poisson equations.

We investigate the modes admitted by both the self-con\-sis\-tent and a
cut-out Mestel disk, the difference being that in the latter, a
part of the matter in the disk is immobilised.  This breaks the
self-similarity and produces a pronouncedly different picture,
both technically and in terms of the disk properties.  The
self-consistent disk is governed by a Cauchy
integral equation, the cut-out disk by an integral equation
that can be treated as a Fredholm equation of the second kind.

In general, our approximation reproduces the results of the previous
works remarkably well, yielding quantities mostly within 5\% of the values reported by Zang and Evans \& Read and thus also
the basic result that in a ``standard''
cut-out disk, only one-armed modes are unstable at the limit of
axisymmetric stability. 
In the self-consistent disk, relatively compact expressions for the kernel
allow an intuitive understanding of most of the properties of neutral
(non-rotating, non-growing) modes there.  We finally show that 
self-consistent Mestel disks do not admit growing or rotating
modes in this sort of stellar-dynamical analysis.

   \end{abstract}

%

\section{Introduction}

It was not long after Kalnajs (\cite{kala}, \cite{kalb},
and references therein) and others had laid out a programme
to perform stability analyses for galactic disks that it was applied
to the Mestel disk by Zang (\cite{1976PhDT........26Z}; henceforth 
cited as Zang) in a Ph.D.~Thesis
supervised by Alar~Toomre.
This disk, characterised by an infinitesimal height and a
globally constant rotation curve, lends itself as a model for
disks that show spiral structure mainly for two reasons.  For one,
its structure is so simple that one of its
distribution functions
has made it into a textbook example (Binney~\&~Tremaine \cite{bintre});  secondly, and probably
more importantly, flat rotation curves seem to be rather prevalent
among (high surface brightness) spiral galaxies 
(e.g., Persic~\&~Salucci \cite{persal}),
and it has long been suspected that flat rotation
curves and strong, ``modal-looking'' spiral structure 
are somehow linked (e.g., Biviano et al.~\cite{biviano}).

One of the drawbacks of the Mestel disk is that the orbits in it cannot be
expressed in elementary functions.  Zang circumvented this problem
essentially through numerical integration and some elegant devices
exploiting the self-similarity of the Mestel disk (i.e., orbits of a
given eccentricity look like scaled copies of each other regardless
of the angular momentum of a particle on them) to curb down on the
computational effort.  While the self-similarity is welcome in
simplifying the calculation, it also implies that a
self-consistent Mestel disk does not possess any scales.  In particular, the
absence of a time scale already leads to the expectation that discrete modes
will be hard to construct.

Zang found that the Mestel disk is a rather bewildering construct.
His formalism leads to a singular integral equation that is quite difficult
to treat numerically. However, the structure of its kernel as
well as physical arguments suggest that there are
no discrete modes apart from rather exotic non-rotating and
non-growing ones, implying that either no growing modes
exist or modes become unstable regardless of their growth
rate and pattern speed. Thus, his investigations concentrated on a variant of
the Mestel disk in which a certain part of the disk is declared
``immobile'', the cut-out disk.  This tampered system is 
furnished with a length scale,
admits discrete modes and becomes treatable with his numeric
technique.

But even the cut-out disk behaves contrary to the preconceptions of the
mid-70ies.  While
Zang would have liked to see a mildly unstable two-armed mode, typically the
only mode that is growing in an axisymmetrically stable disk
is a one-armed mode, and for this, a stability limit cannot
be found.  Even when he made the inner cut-out so steep that
$m=2$ modes emerged in axisymmetrically stable disks in a way that Toomre 
(\cite{1977ARA+A..15..437T}) describes as unrealistic, these one-armed
modes, again in Toomre's words continued
to ``plague'' him.

Despite these somewhat unexpected findings and the fact that is was
only fragmentarily published in Toomre (\cite{1977ARA+A..15..437T})
and Toomre (\cite{toomreswing}), Zang's thesis became a
frequently-cited classic and the Mestel disk developed a life of its
own.  Recently, Evans \& Read (\cite{1998MNRAS.300...83E}, 
\cite{1998MNRAS.300...83E}; see
also Read (\cite{1997PhDT.........1R}; henceforth cited as ERI and ERII) 
generalised
Zang's work to the entire family of physically plausible disks
with surface densities following a power law, and 
Goodman~\&~Evans (\cite{goodev}) once more tackled the issue of rotating modes
the untampered (self-consistent) Mestel disk.

This work set out as an attempt to avoid the numerical integrations
that had to be employed by Zang and ERI by approximating
the orbits and eventually deriving a
closed kernel of the integral equation governing the problem.  Among
the merits of the availability 
of such a kernel is that an analytic expression
for the kernel might allow a decision whether or not the self-consistent
Mestel disk admits rotating modes.
Also, it might facilitate an analysis
of the processes at work at various resonances of the modes in
the cut-out disk in the context of a full disk without having
to take recourse to somewhat daring techniques to extend WKBJ 
approximations across resonances (e.g., Mark \cite{mark}).

The organisation of this paper is as follows:  In section~2, we derive
the orbits and the Hamiltonian of the disk.  Section~3 basically
follows Kalnajs' programme that will in the end yield
an (almost) closed expression for the kernel of
an integral equation governing the modes in the disk.  
To assess the validity of this kernel, we will compare
the results reported in ERII
with ours in section~4.  Finally,
we revisit the issue of modes in the self-consistent disk in section~5.

\section{The Orbits}

The equations of motion in the Mestel disk are
\begin{eqnarray}\ddot
r&=&r\dot\theta^2-1/r
\cr L_z&=&r^2\dot\theta={\rm const}.\label{eqmotmestel}
\end{eqnarray}
Here, $r$ and $\theta$ denote polar coordinates, and $L_z$ is the 
$z$-component of the angular momentum per unit mass.  
Units have been chosen such that
the circular velocity $v_c=2\pi {\rm G}\Sigma_0=1$, where ${\rm G}$ is 
the gravitational constant, and $\Sigma_0$ is the surface density
at a reference radius that can be set to unity because of the self-similarity
of the disk.

We seek closed expressions for the solution of the equations 
(\ref{eqmotmestel}).
Since the exact solutions cannot be expressed in elementary
functions, we need to employ some approximation.  A straightforward
linearisation in configuration space -- the epicyclic approximation --
fails for the Mestel disk, since the Hamiltonian derived from it
does not contain contributions from non-circular motion components.
Therefore we linearise in Fourier space, 
a procedure generally known as the method of
harmonic balance (Bogoliubov~\&~Mitropolski \cite{bogmip}).

Inserting one of the equations (\ref{eqmotmestel}) into the other yields
\begin{equation}\ddot r=L_z^2/r^3-1/r.\label{eqmotmesteln}\end{equation}
We now substitute $r(t)=R_0(1+\zeta\cos(\kappa t))$, making $R_0$ the reference radius
of a given orbit, $\zeta$ a measure 
for its deviation from circularity, and $\kappa$ an
epicyclic frequency.  After this, the first Fourier
coefficients of
(\ref{eqmotmesteln}) with respect to $t$ are
\begin{eqnarray}
a_0&=&{L_z^2\bigl(1+\zeta^2/2\bigr)-R_0^2\bigl(1-\zeta^2\bigr)^2\over R_0^3\bigl(1-\zeta^2\bigr)^{5/2}
}\\
a_1&=&-{2R_0^2\bigl(1-\zeta^2\bigr)^2\bigl(\sqrt{1-\zeta^2}-1\bigr)+3L_z^2\zeta^2\over\zeta R_0^3\bigl(1-\zeta^2\bigr)^{5/2}}\\
b_1&=&0,\label{eqmotfourier}\end{eqnarray}
where $a_i$ and $b_i$ denote the cosine and sine coefficients, respectively.

Dropping all higher Fourier coefficients and substituting the result back
into (\ref{eqmotmesteln}), one finds by comparing the terms independent of $t$
\begin{equation}
L_z={\sqrt2R_0(1-\zeta^2)\over \sqrt{\zeta^2+2}},\label{Lzdeflong}
\end{equation}
and by comparing terms proportional to $\cos(\kappa t)$
\begin{equation}
\kappa= {\sqrt2\over \zeta R_0}\sqrt{1-{2{\sqrt{1-\zeta^2}\over\zeta^2+2}}}.
\label{kappadeflong}
\end{equation}
The circular frequency of the guiding centre $\Omega$ results from a 
Fourier transformation of the second equation in (\ref{eqmotmestel})  with
(\ref{Lzdeflong}) inserted and is given by
\begin{equation}
\Omega={L_z\over R_0^2\left(1-\zeta^2\right)^{3/2}}.\label{Omegadeflong}
\end{equation}

We can now derive the Hamiltonian for a mass point in the
Mestel disk in action-angle coordinates by simply inserting (\ref{Lzdeflong}),
(\ref{kappadeflong}), and (\ref{Omegadeflong}) into an Hamiltonian written
in Cartesian coordinates.  Assuming orbits not too far from circularity,
one can linearise with respect to $\zeta$.  After again dropping
all Fourier components of order 2~or higher, one is left with
\begin{equation}
H(J,L_z)={1\over2}+\ln(L_z)+{\sqrt2 J\over L_z},\label{linhamil}
\end{equation}
where $J=\kappa\zeta^2R_0^2/2$ is the energy in the epicyclic motion divided
by $\kappa$ and thus the second integral.  This clearly is the simplest
Hamiltonian that could have anything to do with a Mestel disk and indeed
the Hamiltonian that would result from a classic epicyclic approximation
if it worked.  As a matter of fact, at this point
one has $\kappa=\sqrt2/R_0$ and 
$\Omega=1/R_0$, reproducing the result of an epicyclic analysis.
This Hamiltonian has already been used by Collett 
et.~al.~(\cite{1997MNRAS.285...49C}).

The set of approximated action-angle coordinates $L_z$, $J$,
$w_1:=\kappa t$,
$w_2:=\Omega t$ is correct to first order in $\zeta$, 
and the canonical equations are correct to first order in $\zeta$ as well.
As is demonstrated in Demleitner (\cite{diss}), the 
Jacobian of the transformation
between Cartesian coordinates and this set is $1+O(\zeta^2)$, showing that the
transformation is indeed canonical to the required order.

Because of the two approximations (only keeping the lowest Fourier coefficients
and linearising with respect to $\zeta$), 
it is not \emph{a priori} clear that the disk described by this transformation
and the resulting Hamiltonian (\ref{linhamil}) does indeed behave like
a Mestel disk.  It is one of the purposes of this work to show \emph{a
posteriori}, at is were, that this is the case.

\section{Derivation of the matrix equation}

With the Hamiltonian and the transformation equations, we can now follow 
the programme outlined in the series of papers by Kalnajs cited
above.  The first step is to establish a base for
the potential perturbation.  We choose
\begin{eqnarray}
\phi_{\alpha,m}&=&L_z^{i\alpha-1/2}{\rm e}^{imw_2-i\omega t}\cr&&
{\rm e}^{\left(i\sqrt{\sqrt2J/L_z}\bigl(\alpha\cos(w_1)-
\sqrt2 m \sin(w_1)\bigr)\right).}\label{potbaslin}
\end{eqnarray}
Here, $m$ is the number of circumferential maxima of the spiral
pattern (the number of arms), and $\alpha$ parametrises how 
closely the spiral is wound, where a spiral with larger $\alpha$ is more 
trailing provided that $\omega>0$.  This expression basically is the
well-known logarithmic spiral ${\rm e}^{i\alpha\ln(r)+i m\theta-i\omega
t}$ transformed into our action-angle coordinates with harmonics
in the angle coordinates higher than one disregarded.  The factor
$L_z^{-1/2}$ was added to ensure definiteness in the integration
of the angular momentum.  This analogy motivates following ERI in naming 
$\alpha$ 
the logarithmic wave number and $m$ the angular harmonic number.  We assume
that all potential perturbations can be represented as expansions of
the form
\begin{equation}
\sum_{m=0}^\infty\int_{-\infty}^\infty\Phi_m(\alpha)
\phi_{\alpha,m}(w_1,w_2,J,L_z)\,d\alpha.\label{potexpansion}
\end{equation}

The next step in Kalnajs' programme is to solve the linearised Boltzmann equation
\begin{equation}
{\partial f_1\over\partial t}+[f_1,H]=-[f_0,\phi_{\alpha,m}],\label{linbol}
\end{equation}
written with Poisson brackets $[\,.\,,.\,]$ 
for an elementary perturbations $\phi_{\alpha,m}(w_1,w_2,J,L_z)$ 
of the form (\ref{potbaslin}).
Here, $f_1$ is the unknown perturbation of the distribution
function, and $f_0$ an equilibrium distribution function; for $f_0$,
we use the distribution function given by Binney~\&~Tremaine 
(\cite{bintre}), which, in our units and coordinates, and after applying
Stirling's formula, is
\begin{eqnarray}
f_0(L_z,J)&=&{F\over L_z}\exp\left(-{\sqrt2J\over
\sigma^2 L_z}\right)\quad{\rm with}\nonumber\\
F&=&{\hbox to 0pt{$\sqrt{1-2\sigma^2}^{1-1/\sigma^2}$\hss}\phantom{\sqrt{1-2\sigma^2}1}
\over 2\sqrt2{\rm e}\pi^2\sigma^2{\rm G}.}
\label{Fdefcooked}
\end{eqnarray}

Expanding the Poisson brackets in (\ref{linbol}) yields 
\begin{eqnarray}
{\partial f_1\over\partial t}+&\omit&\!\!
{\sqrt2\over L_z}{\partial f_1\over\partial w_1}+
{1\over L_z}{\partial f_1\over\partial w_2}=
-i{F
L_z^{i\alpha}{\rm e}^{imw_2-i\omega t}\over\sigma^2L_z^{5/2}}\cr&\omit&
\left(X(w_1)-
m\sigma^2\right)
\exp\left(iX(w_1)-{\sqrt2J\over\sigma^2L_z}\right)
\label{bolexp}
\end{eqnarray}
with an auxiliary function 
$X(w_1)=\sqrt{\sqrt2J/L_z}\Bigl(\alpha\cos(w_1)-\sqrt2m\sin(w_1)\Bigr)$.
On the right-hand side of this equation, a term of order $\zeta^2$ has
been neglected.

To solve this equation, we perform a separation of variables by
\begin{equation}
f_1(w_1,w_2,t)=g_1(w_1){\rm e}^{imw_2-i\omega t}
\label{sepans}
\end{equation}
and arrive
at an inhomogeneous ordinary differential equation, the homogeneous
solution of which is 
\begin{equation}
\tilde g_1(w_1)={\rm e}^{i(\omega L_z-m)w_1/\sqrt2}.
\label{hsol}
\end{equation}
The general solution can be obtained via a variation
of constants, where the integration constant is fixed by the 
condition that the distribution function must be $2\pi$-periodic,
$g_1(w_1)=g_1(w_1+2\pi)$.

Before writing down the result, we note that in (\ref{bolexp}) the terms
with $w_1$ in the exponent and its coefficient are quite similar
to each other.  This invites a partial integration in the resulting
expression, which after some algebra leads to
\begin{eqnarray}
g_1(w_1)&=&C\sqrt{L_z\over\sqrt2J}
\exp\left(-{\sqrt2J\over\sigma^2L_z}\right)
\Biggl(
i\,\exp\bigl(iX(w_1)\bigr)\cr
&&-{1\over {\rm e}^{-2\pi i\eta}-1}
\left(\sqrt2\eta+m\sigma^2\right)\cr
&&\int_0^{2\pi}\exp\bigl(iX(w_1+w_1')-i\eta w_1'\bigr)\,dw'_1\Biggr).
\label{solc}
\end{eqnarray}
In this expression, we have abbreviated $\eta=(\omega L_z-m)/\sqrt2$.

Substituting (\ref{solc}) into (\ref{sepans}) yields the perturbation
of the distribution function under an elementary potential perturbation 
$\phi_{\alpha,m}$.  Since (\ref{linbol}) is linear in $\phi_{\alpha,m}$, the
response
to an arbitrary perturbation $\Phi_1^{\rm imposed}$ 
of the form (\ref{potexpansion})
can immediately be computed by summing over all $\alpha$ and
$m$.  Thus, the solution of (\ref{linbol}) can be written as
\begin{equation}
f_1^{\rm response}={\cal L}\Phi^{\rm imposed}_1\label{opeq}
\end{equation}
with a linear integral operator $\cal L$.

Further pursuing Kalnajs' programme, we attempt to find a linear combination
of elementary perturbations such that $f_1^{\rm response}$
causes the $\Phi^{\rm imposed}_1$ we started with.
To do this, the Poisson equation has to be solved simultaneously to
the Boltzmann equation.  Since $\cal L$ and the Laplacian occurring
in the Poisson equation
do not commute, their eigenfunctions are different, and a
straightforward attempt to solve the two equations
simultaneously will result in an integro-differential equation
that will be very hard to solve indeed.

To avoid this, one converts the system of equations to a matrix equation.
The idea here is to compute potential-density pairs, i.e., two sets of
mutually orthogonal basis functions $\phi_{\alpha,m}$ and $\mu_{\alpha,m}$
for each of the perturbations in density and 
potential where each pair  $\phi_{\alpha,m}, \mu_{\alpha,m}$ 
satisfies the Poisson equation. 
When one expands the perturbations in the potential and the
surface density
in these functions, the Poisson equation simply reads $\Phi_m(\alpha)=
M_m(\alpha)$ in the coefficients $\Phi_m(\alpha)$ and $M_m(\alpha)$ of
these expansions.

The basis $\phi_{\alpha,m}$ is already fixed
by (\ref{potbaslin}), and the corresponding density base function
can be found using the scheme of Clutton-Brock (\cite{clubro}).  
For our calculation, however,
the only important property of the potential-density pair
the scalar
product of $\phi_{\alpha,m}$ and $\mu_{\alpha,m}$.  This scalar product
is of course invariant under canonical transformations, and to the the required
order Clutton-Brock's result 
\begin{equation}
\int_0^{2\pi}\!\!\!d\theta\int_0^\infty\!\!\!r\,dr\,\phi_{\alpha,m}^\ast
\mu_{\alpha',m'}=4\pi^2K_m(\alpha)\delta_{m,m'}\delta(\alpha-\alpha')
\label{basscalprod}
\end{equation}
for logarithmic spirals holds.  Here,
\begin{equation}
K_m(\alpha)={
\left|\Gamma\left({3\over4}+{1\over2}m+{1\over2}i\alpha\right)\right|^2\over
\pi{\rm G}\left|\Gamma\left({1\over4}+{1\over2}m+{1\over2}i\alpha\right)\right|^2}
\label{Kdef}
\end{equation}
is a quantity closely related to ERI's Kalnajs factor.

After expanding the potential and surface density perturbations in (\ref{opeq}),
one can project both sides on $\phi_{\alpha',m'}$. Exploiting
(\ref{basscalprod}) and inserting the (now trivial) Poisson equation,
(\ref{opeq}) takes the form
\begin{eqnarray}
4\pi^2&\omit&\!\!
K_{m'}(\alpha')\Phi^{\rm response}_{m'}(\alpha')=
\int_{-\infty}^\infty\!d\alpha
\int_0^\infty\!dL_z\int_0^{\infty}\!dJ
\cr&\omit&\displaystyle\int_0^{2\pi}\!dw_1
\sum_{m=0}^\infty\int_0^{2\pi}\!dw_2\,
\Phi^{\rm imposed}_m(\alpha)({\cal L}\phi_{\alpha,m})
\phi^\ast_{\alpha',m'}.
\label{mateq}
\end{eqnarray}
Equating $\Phi^{\rm response}$ and $\Phi^{\rm imposed}$, an integral
equation results the solutions of which, if they exist, are self-consistent
modes.

The integrand of (\ref{mateq}) can be written as a sum $I_1+I_2$ of
a part without a further integral,
\begin{eqnarray}
I_1&=&\Phi_{\alpha,m}{FL_z^{i(\alpha-\alpha')-2}
\over4\pi^2\sigma^2K_{m'}(\alpha')}
{\rm e}^{i(m-m')w_2-{\sqrt2J\over\sigma^2L_z}}\cr
&&{\rm e}^{i\sqrt{\sqrt2J\over
L_z}\bigl((\alpha-\alpha')\cos(w_1)-
\sqrt2(m-m')\sin(w_1)\bigr)},
\label{intgrone}
\end{eqnarray}
and one in which a further integration over the auxiliary angle
$w_1'$ has to be performed,
\begin{eqnarray}
I_2&=&\int_0^{2\pi}dw_1'\,i\Phi_{\alpha,m}{\sqrt2F
L_z^{i(\alpha-\alpha')-2} \bigl(m\sigma^2+\sqrt2\eta\bigr)\over
8\pi^2\sigma^2K_{m'}(\alpha') ({\rm e}^{-2\pi i\eta}-1)}
\cr&&{\rm e}^{i(m-m')w_2-i\eta w_1'-{\sqrt2J\over
\sigma^2L_z}+\sqrt2\,m'\sin(w_1)}\cr&&{\rm e}^{i\sqrt{\sqrt2J\over L_z}
\left(\alpha\cos(w_1'+w_1)-\alpha'\cos(w_1)
-\sqrt2\,m\sin(w_1'+w_1)\right)}.
\label{intgrtwo}
\end{eqnarray}

\begin{figure}
\resizebox{\hsize}{!}{\includegraphics{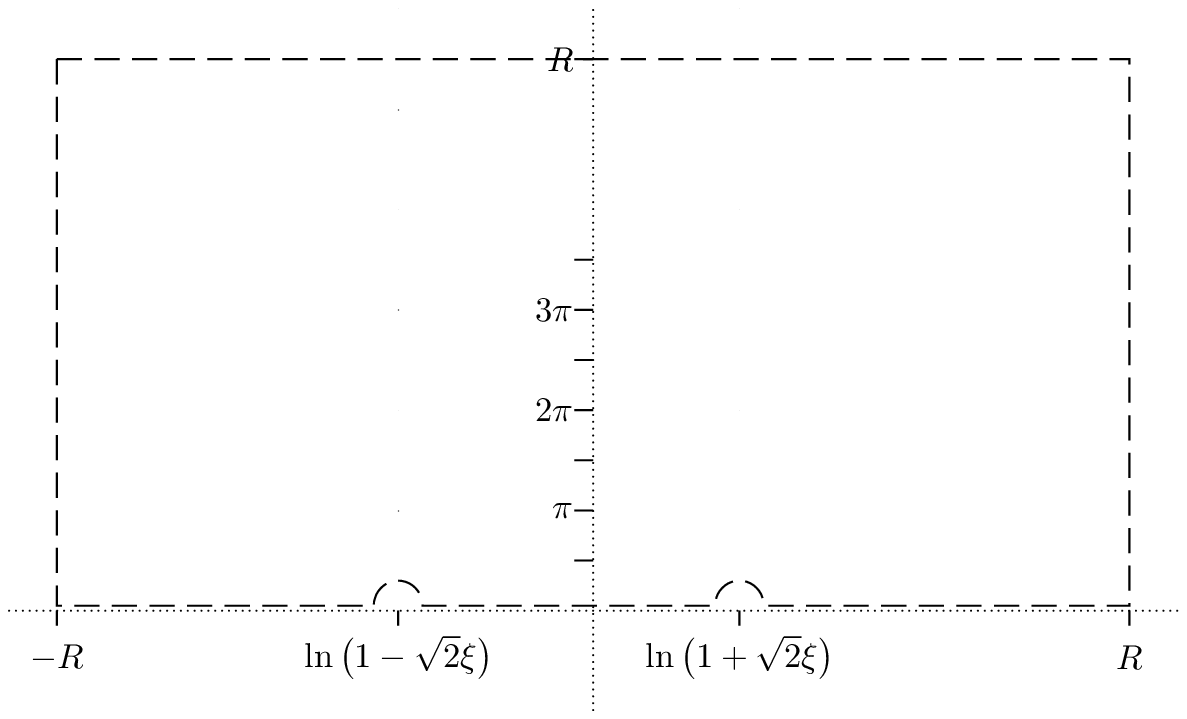}}
\caption{Poles of the integrand of (\ref{stointtwon}) and a sketch
of the integration path.  The diamonds in the left column mark
the positions of the poles at $\Re(u)=\ln(1-\sqrt2\xi)$ 
in the case $\xi>1/\sqrt2$.}
\label{intpath}
\end{figure}

The quadratures over $w_2$, $w_1$, and $J$, carried out in that
order, are rather
straightforward.  Details can be found in Demleitner (\cite{diss}).
When integrating over $w_1'$ one has to evaluate an expression
of the type 
\begin{equation}
\int_0^{2\pi}{\rm e}^{i\tau w_1'}{\rm e}^{-\nu\cos(v-w_1')}\,dw_1',
\label{woneprime}
\end{equation}
which can be done by expanding the second factor of the integrand
into modified Bessel functions ${\rm I}_k$ using
\begin{equation}
{\rm e}^{-\nu\cos(v-w_1')}={\rm I}_0(\nu)+2\sum_{k=1}^\infty{\rm
I}_k(\nu)\cos\bigl(k(v-w_1')\bigr)
\label{woneprimetrick}
\end{equation}
(Abramowitz~\&~Stegun \cite{abrsteg}).
Unfortunately, the series introduced by this operation does not in
general collapse and will be present in the kernel.

The integration over the angular momentum is the most involved one,
resembling the calculations to compute the angular momentum function in
Zang's formalism.
For the summands of the series resulting from the $w_1'$
integration, one has to treat integrals
of the structure
\begin{equation}
\int_{-\infty}^{\infty}{{\rm e}^{(i\gamma+\lambda+1) u}\over
2\,\bigl(2\xi^2-(1-{\rm e}^u)^2\bigr)}\,du,\label{stointtwon}
\end{equation}
where 
\begin{equation}
u=\ln(\omega L_z/m)
\label{udef}
\end{equation}
(note that this substitution only works for
nonzero $m$ and $\omega$), $\gamma=\alpha-\alpha'$ and $\xi=k/m$. The integer
$k$ is 
a summation index varying between $0$ and $\infty$, and the complete
integral requires this expression to be evaluated for
$\lambda\in\{-1,0,1\}$.

We will compute this integral using the residue theorem, and thus
the first step is to find the zeroes of the denominator of the integrand.
These are at 
\begin{equation}
u^{(1,2)}_j=\ln(1\pm\sqrt2\xi)+2\pi ij
\label{stwointpoles}
\end{equation}
for integer $j$.  These zeroes are well-defined, since $\xi$, as
the ratio of two integers, can never be equal to $1/\sqrt2$.
Let for the moment $\xi<1/\sqrt 2$ and $\gamma>0$.

There are three more or less technical problems in the application of the
residue theorem to this integral.  The first is that at least one pole is
on the real axis and hence on the integration contour.  The tempting idea
to take the (existing) Cauchy principal value leads to erroneous results.
As Landau (\cite{landau}) noted in a plasma-physical context, the correct
reasoning is that if a system supports a non-growing perturbation at
present and started in equilibrium, the perturbation must have been a
growing one at some time in the past.  Since it has evolved continuously,
the integration contours of the problem must have been continuously
deformed as well and thus cannot have jumped across poles.  A growing
perturbation corresponds to an $\omega$ with a positive imaginary part,
which with the substitution (\ref{udef}) means that the integration
contour moves into the upper half-plane.  Therefore, as $\Im\omega\to0$,
one has to leave the zeroes on the real axis out of the
contour, as is sketched in Fig.~\ref{intpath}.

The second problem is that as the size of the
integration path $R\to\infty$ (cf.~Fig.~\ref{intpath})
one cannot continuously deform the contour since the poles are
in the way, and the same poles make it hard to establish that the
contribution of the line at $\Im u=iR$ vanishes.  
The standard technique to avoid this is to write down the series of 
residues and show that integral over the difference between this series and
the integrand vanishes as $R\to\infty$.  Since this differences has no poles, the contour
can be readily deformed.

The last problem is that for $\lambda=\pm1$ the contributions at
$\Re u=\pm R$ do not necessarily vanish.  Indeed, one finds that 
these ``border terms''
\begin{equation}
=\cases{-{i\over 4\gamma\xi^2-2}\lim_{x\to-\infty}
{{\rm e}^{i\gamma x}\over\gamma};&$\lambda=-1$, $u\to-\infty$\cr
-{i\over2}\lim_{x\to-\infty}{{\rm e}^{i\gamma x}\over\gamma};&$\lambda=1$, $u\to\infty$}
\label{bordterms}
\end{equation}

Of course, the limits occurring in these expressions do not exist.
However, it should be remembered that this integral is being evaluated
as part of the kernel of an integral equation.  Therefore, 
the limits can (and have to) be taken in the sense of a distribution limit,
$\lim_{x\to\pm\infty}{\rm e}^{i\gamma x}/\gamma=\pm
i\pi\delta(\gamma)$.

With these tools, one can go about evaluating the residues of the 
integrand, which can easily be done using ${\rm res}(f,z)=
\lim_{u\to z}f(u)(u-z)$ and an application of de l'Hospital's rule.
We will give the end result below.

A few words about the other cases are in order.  First, if 
$\xi>1/\sqrt2$, the poles at $u_j^{(1)}$ move away from the
real axis (cf.~Fig.~\ref{intpath}).
One can keep the expression resulting
for $\xi<1/\sqrt2$ by choosing an appropriate branch of the logarithm 
in the result.  Basically, one has to see to it that the $u_j^{(1)}$ in
Fig.~\ref{intpath} ``shift down'' rather than ``up'' by $i\pi$.  To ensure
this, one sets $\ln(-1)=-i\pi$, and with $\ln(1)=0$ a natural
branch cut is along the positive imaginary axis, so that ${i\pi\over2}\geq
\ln({\rm e}^{i\phi})>-{3\pi i/2}$.

If $\gamma<0$, one has to integrate through the lower half-plane, and
consequently the poles on the real axis are added into the residue sum.
The resulting expression is, of course, identical to the one for $\gamma>0$.

After carrying out all integrations and the summation of $m$ (which
is trivial because the integration over $w_2$ contributes a $\delta_{m,m'}$, thus
decoupling different angular harmonics), (\ref{mateq}) takes the form
\begin{equation}
\Phi_{m'}(\alpha')={\cal F}(\alpha')\Phi_{m'}(\alpha')+\int_{-\infty}^\infty
{\cal K}(\alpha,\alpha')\Phi_{m'}(\alpha)\,d\alpha.
\label{skernform}
\end{equation}
The summand with $\cal F$ originates from the terms with 
$\delta$-functions and computes to
\begin{eqnarray}
{\cal F}(\alpha')&=&{\sqrt{1-2\sigma^2}^{1-1/\sigma^2}\over 
2\sqrt2\sigma^2}{
\left|\Gamma\left({1\over4}+{1\over2}m+{1\over2}i\alpha'\right)\right|^2
\over
\left|\Gamma\left({3\over4}+{1\over2}m+{1\over2}i\alpha'\right)\right|^2
}
\Biggl(1-\cr&&(1+\sigma^2)
{\rm e}^{-\sigma^2(\alpha'^2+2m'^2)/2}
\biggl({\rm I}_0\left({\sigma^2\over2}
(\alpha'^2+2m'^2)\right)\cr&&
+\sum_{k=1}^\infty {2m'^2\over m'^2-2k^2}\,{\rm I}_k\left({\sigma^2\over2}({\alpha'}^2+2{m'}^2)\right)
\biggr)\Biggr).
\label{skernfdef}
\end{eqnarray}

The kernel itself becomes
\begin{eqnarray}
{\cal K}&\omit&\!\!(\alpha,\alpha')=-i
{\omega^{-i(\alpha-\alpha')}\sqrt{1-2\sigma^2}^{1-1/\sigma^2}\over
2\sigma^2{\rm e}^{\sigma^2(\alpha^2+\alpha'^2+4m'^2)/4+1}
\bigl({\rm e}^{2\pi(\alpha-\alpha')}-1\bigr)}\cr&\omit&
{\left|\Gamma\left({1\over4}+{1\over2}m+{1\over2}i\alpha'\right)\right|^2
\over
\left|\Gamma\left({3\over4}+{1\over2}m+{1\over2}i\alpha'\right)\right|^2
}
\Biggl(m'^{i(\alpha-\alpha')}\sigma^2
{\rm I}_0(\tau)
\cr&\omit&-\sum_{k=1}^\infty
{\rm I}_k(\tau)
\biggl((\sqrt2k-m'\sigma^2)\Lambda^k
{\rm e}^{(i(\alpha-\alpha')-1)\ln(m'-\sqrt2k)}
\cr&\omit&-(\sqrt2k+m'\sigma^2)\Lambda^{-k}
{\rm e}^{(i(\alpha-\alpha')-1)\ln(m'+\sqrt2k)}\Biggr),
\label{skernkdef}
\end{eqnarray}
where we have abbreviated
\begin{eqnarray}
\tau&=&{\sigma^2\over2}
\sqrt{(\alpha^2+2m'^2)(\alpha'^2+2m'^2)}\cr
\Lambda&=&{\alpha\alpha'+\sqrt2m'(\sqrt2m'+i\bigl(\alpha-\alpha')\bigr)\over
\sqrt{(2m'^2+\alpha^2)(2m'^2+\alpha'^2)}}.\label{kernabbrevs}
\end{eqnarray}

Formally, (\ref{skernform}) is an integral equation of the second kind.
However, $\cal K$ is not compact -- the infinite integration range
and the singularity on the $\alpha=\alpha'$ diagonal prevent this --,
and thus the solution theory for it is significantly different from
what one has in the Fredholm case.  We will return to this issue
later.  For now, let us just follow Zang in noting that 
$\omega$ factors out of
the kernel, and thus the
value of $\omega$ is irrelevant for the solvability of this
equation.

The special case of $\omega=0$ has not been treated so far.  It is
interesting for two reasons:  For one, when searching for the limit
of axisymmetric
stability, it suffices to consider $\omega=0$, since axisymmetric
patterns cannot rotate anyway and considering growth is not 
necessary for establishing the stability margin, and for two, ``neutral'' modes
(for which $\omega=0$ even when $m\neq0$) 
were the only modes the existence of which ERII and Zang 
report for the self-consistent disk.

The expression (\ref{skernkdef}) cannot be taken to the limit $\omega\to0$, because 
$\omega^{i(\alpha-\beta)}$ in general has an essential singularity
at $\omega=0$, which according to Picard's theorem implies that the
image of an arbitrarily small neighbourhood of 0 is the entire complex
plane with the possible exception of one point.  Instead, one has to
substitute $\omega=0$ before the integration over $L_z$.  This
simplifies the terms quite significantly, resulting in
\begin{equation}
{\cal K}_0(\alpha')=2{\cal F}(\alpha')
\label{somzerokernel}
\end{equation}
The integral equation has now collapsed into the algebraic equation
\begin{eqnarray}
\Phi_{m'}(\alpha')&=&{\cal K}_0(\alpha')\Phi_{m'}(\alpha'),
\quad\hbox{or}\cr
0&=&{\cal K}_0(\alpha')-1,
\label{somzerointeq}
\end{eqnarray}
which clearly simplifies solving the equation by at least an order of
magnitude.

\section{The cut-out disk}

The Mestel disk has three rather bizarre properties:
Its mass is infinite, the surface density is not bounded in the centre,
and it does not possess a natural time-scale.  All these shortcomings
can be cured by a relatively simple measure that was already employed by
Zang.  The idea is to immobilise a radially varying part of the mass,
declaring it as, say, halo or bulge.  Thus one keeps the
simple structure of the disk, but cuts the singularity in the centre and
possibly the long mass tail out to infinity as far as the Boltzmann
equation is concerned.  Also, the cut-out will break the
self-similarity and thus introduces a time scale.

As an added benefit, the ratio of active to total mass
and thus (at radially constant velocity dispersion) 
Toomre's stability parameter $Q$ (Toomre
\cite{toomreq}) becomes a
function of the radius, opening the possibility to introduce a
$Q$-barrier in the disk.  In such a $Q$-barrier, incoming waves are refracted
away from the centre and thereby evade their absorption at the inner
Lindblad resonance.  In common semi-phenomenological theories of spiral structure
(Bertin et.~al.~\cite{bertin}), this
$Q$-barrier serves as a part of a feedback loop, to be closed by, e.g.,
the corotation at which swing amplification (Toomre
\cite{toomreswing}) or related mechanisms would not only provide
reflection but even amplification of wave packets.  

In the shearing sheet
(Goldreich~\&~Lynden-Bell \cite{1965MNRAS.130..125G}, 
Julian~\&~Toomre \cite{1966ApJ...146..810J}), it is possible to show
rigorously that the introduction of something quite similar to a $Q$-barrier
leads to the emergence of stationary modes where no
such modes are present in the classic border-less analyses (Fuchs 
\cite{fuchspre}).

Technically, the immobilisation of the mass is done by multiplying
the distribution function with a cut-out function $H(r)$, 
with $0\leq H(r)\leq1$, while keeping the Hamiltonian and thus the
rotation curve unchanged.
Since the main objective of our investigation of the cut-out disk was
an assessment of the quality of our approximation, we used the cut-out function
used by ERI (who in turn basically used what Zang had suggested).
In our case, the introduction of the cut-out has the effect of eliminating
the border terms in the $L_z$ integration, but only if one uses Zang's
doubly cut-out disk (i.e., the cut-out function tends to zero for
both $r\to0$ and $r\to\infty$).  Zang's family of cut-out functions is
parametrised by two positive integers $M$ and $N$, the cut-out indices, 
describing the steepness
of the cut-outs in the outskirts and centre of the disk, respectively.
Since both of these integers appear as the order of poles 
of the integrand in the $L_z$ integration, and the residues at these
poles get very messy indeed as their order increases, we settled
for the case $M=N=2$, resulting in a cut-out function
\begin{equation}
H(L_z)={L_z^2L_c^2\over(L_z^2+L_0^2)(L^2_z+L_c^2)},
\label{readcutout}
\end{equation}
where $L_0$ and $L_c$ are two parameters qualifying the locations
of the inner and outer cut-outs;  following ERII, we usually use
$L_0=1$ and $L_c=10$.

It should be noted that for further analytic exploration of the cut-out
disk, this cut-out is not favourable, since it still leads to a very
cumbersome kernel.  A cut-out of 
$H(L_z)=2L_zL_0/(L_0^2+L_z^2)$ results in a much more compact kernel
and reproduces the overall behaviour the cut-out disk
(due to its lower active surface density,
the resulting disk has an even higher stability than one with Zang's cut-out,
though).

Since the cut-out function (\ref{readcutout}) only depends on $L_z$,
most of the calculations done for the self-consistent disk just
carry over.  The integration over $L_z$ becomes somewhat easier,
since there are no border terms, but this advantage is more
than offset by the sheer length of the terms and the occurrence of higher-order poles.  The resulting
integral equation has an ``almost'' compact kernel, with no singularity
on the diagonal and a relatively sharp decline of the kernel as
$\alpha$ and $\alpha'$ approach infinity.  Consequently, ERI's naive approach of
simply Fredholm-discretizing the integral equation and cutting
off
at some appropriate $\alpha$ still works well.  The only difference
to ERI's method of solving this integral equation was that
we found a spline-based quadrature to converge faster than
the integration scheme proposed by them.

Skipping the details -- which can be found in Demleitner (\cite{diss}) --,
let us briefly compare the stability of disks described with our approximation 
with Read's findings.  Since ERII mostly report results for
singly cut-out disks which are not favourable for an anaylsis using
our tools, we give the values from Read's (\cite{1997PhDT.........1R}) more
comprehensive tables. 
Our disks become unstable to axisymmetric 
($m=0$) perturbations at $\sigma_{\rm min}=0.274$;  
Read only gives values for
$L_c=100$, where we find $\sigma_{\rm min}=0.349$ versus Read's 
$\sigma_{\rm min}=0.327$. This agreement to within roughly 5\% is remarkable
considering the relatively high temperature of this disk.  The
higher stability of the disk with lower $L_c$ is mostly due to its lower
active surface density.

The one-armed ($m=1$) modes investigated in ERII 
behave somewhat erratic in that at
vanishing growth rate and $\Re \omega\to0$ in that there seem to be modes for
almost arbitrary $\sigma$.  That, however, is a consequence of the fact
that ERII performed these computations without an outer cut-out.  Introducing
an outer cut-out, the disk seems to become stable at $\sigma=0.57$, although
clearly at that point the assumption of a razor-thin disk becomes
quite unrealistic and $v_c\gg\sigma$ is severely violated.

In the -- at least historically -- most interesting case of two-armed
($m=2$) modes, our disks become unstable at $\sigma_{\rm min}=0.214$ with a pattern
speed $\Omega_p=\omega/m=0.548$, whereas Read finds $\sigma_{\rm min}=0.205$
and $\Omega_p=\omega/m=0.542$.  Thus, not only is the qualitative result
that the $M=N=2$ cut-out disk is stable with respect to two-armed perturbations
reproduced, the quantitative agreement again is far better than one
might expect given the rather severe assumptions that go into our
model.

Read (\cite{1997PhDT.........1R}) does not give stability limits for the $m=3$ and $m=4$ perturbations
in doubly cut-out disks, but a comparison of the eigenvalue plots
leads to the expectation that our results -- $\sigma_{\rm min}=0.141$ at
$\Omega_p=0.61$ for $m=3$ and $\sigma_{\rm min}=0.124$ at $\Omega_p=0.588$ for
$m=4$
-- would again reproduce the results of the more exact calculations quite
well.  The $m=4$ stability limit has to be taken with a grain of
salt, since our formalism reproduces modes at high 
velocity dispersion as $\omega\to0$ in accordance with Read's
findings.  It is remarkable, though, that these modes are not quenched by
an outer cut-out like their $m=0$ brethren.

The properties of the modes computed from the kernels are, by and
large, identical to those discussed by ERII, and our model even
shows more subtle features like the rapid overtaking of eigenvalues
when $m=4$.  

As to comparison to simulations, the fact that 
the expressions get very unwieldy for larger $M$ and $N$
is somewhat unfortunate.
Sellwood~\&~Evans (2001) point out that for the global mode to dominate
over particle noise in simulations, it is highly desirable to maintain
a high amplification, implying steep cut-outs and thus relatively
large cut-out indices.  The values chosen by Sellwood~\&~Evans for their
``standard'' disk, $N=4$ and $M=6$, are beyond the reasonable range of
applicability for our approach. However, they compute a doubly cut-out model 
that does not possess any global modes at $Q=1$ (remember that $Q$ refers to
the axisymmetric stability of the self-consistent disk) and state
that the $N=1$ and $M=2$ disk with $L_0=1$
becomes stable with respect to $m=1$
perturbations below $L_c\approx 5$.  The rotation curve they use
in this model is falling but only very slightly so, so that 
a disk with a completely flat rotation curve should behave analogously.
Indeed, ignoring the ``rogue'' eigenvalues
(Demleitner 2000) -- that become relevant in this model --, we find
that even the $m=1$ mode gets quenched below $L_c=4.7$.

Summarising, our approximation does well in reproducing the results from
the previous calculations for the cut-out disks.
For now, the value of this insight is rather limited, since one
still has to take recourse to numerical integration to solve the
integral equation, and the numerics are computationally
about as expensive as those in Zang's formalism, since the expression for the kernel 
is very
large.  Still, using a cut-out more suitable to our formalism
results in more compact and manageable expressions 
that might facilitate a closer look
at the interplay between the $Q$-barrier, the various 
resonances, and density waves in a full-disk analytical model.

\section{The self-consistent disk}

The axisymmetric stability of the self-consistent disk can be determined
by solving the algebraic equation (\ref{somzerointeq}) with
a kernel still somewhat simpler than (\ref{somzerokernel}) that is the result of
setting $m=0$ in (\ref{intgrone}) and (\ref{intgrtwo}), yielding
$\sigma_{\rm min}=0.3860$ with $\alpha=3.51$, within 2\% of 
Read's (\cite{1997PhDT.........1R})
values of $\sigma_{\rm min}=0.3780$ and $\alpha=3.46$.

Since the integral equation describing the self-consistent
disk for non-axisymmetric modes is highly singular,
Zang's formalism could not provide solutions for the full problem.
However, it did yield a family of rather exotic modes
characterised by $\omega=0$ and consisting of a single logarithmic
spiral -- the neutral modes.  It should be noted that these single logarithmic
spirals are scale-free, a property that sets them apart from from any
other disturbance of the form (\ref{potexpansion}).
In our formalism, they can 
be computed by solving the algebraic equation
(\ref{somzerointeq}) with the kernel (\ref{somzerokernel}).  The resulting
behaviour again agrees very well with the findings of ERII and Zang and
has to be called rather peculiar: One-armed perturbations cannot be stabilised
at all, two-armed modes set in below $\sigma=0.155$ (or $\sigma=0.149$
according to Read (\cite{1997PhDT.........1R})), three-armed ones are extremely 
stable and do not set in
down to at least $\sigma=0.05$, whereas four-armed neutral modes again
cannot be stabilised at all.

The neutral modes come in pairs, i.e., if a mode
with logarithmic wave number $\alpha$ exists, so does one with $-\alpha$.
This is a consequence of the anti-spiral theorem (Lynden-Bell~\&~Ostriker 
\cite{antispir}), since these neutral modes are not limits
of growing modes.  The marginal modes in the cut-out disk and possibly
existing modes with nonzero $\omega$ in the self-consistent disk, on the
other hand, are limits of growing modes by the application of
Landau's rule and thus not subject to the anti-spiral theorem.

ERII, from their perspective of analysing the entire family of physically
viable power-law disks, briefly hypothesised that the stability properties
might be related to the emergence of closed orbits of a given symmetry;
these would, according to her reasoning, absorb power from perturbations
and stabilise the disks when present.  In our model, no closed orbits
at all exist, since the ratio of epicyclic and orbital frequencies
is fixed and irrational.  As the model reproduces
the stability properties found by ERII nevertheless, one has to look
for a different explanation.

Turning to (\ref{somzerokernel}), one sees that the
contribution from the terms except the one with the sum is fairly
independent of $m$, whereas the sum does significantly change its
value with $m$.  The term to watch here is the $m'^2/(m'^2-2k^2)$. 
When $m'\approx\sqrt2k$, the corresponding summand is
much larger than its neighbours since the ratio $I_k(x)$ and $I_{k+1}(x)$
basically is $k$ (plus some constant) and everything else does not
depend on $k$; when $m$ is not too large, that ``resonant'' 
summand will dominate
the value of the entire series.
The sign of this resonant summand is determined by the side relative
to the nearest $k$ on which
$m'-\sqrt2k$ reaches its minimum -- for the $m=3$ series, the near-resonance 
is at $k=2$ and thus one has a positive contribution with
$m'-\sqrt2k=0.17$
that enters into the kernel negatively (enhancing stability), whereas
the resonance for the $m=4$ occurs for $m'-\sqrt2k=-0.24$ and
thus destabilises the disk.  The next azimuthal harmonic that has an
$m$ close to a $\sqrt2k$ is 7, and it is again negative; not
surprisingly, the $m=7$ harmonic is quite unstable as well.

Unfortunately, $k$ has no immediate meaning in physical terms,
although (\ref{solc}),
suggests that the $k$-th terms of the series in (\ref{somzerokernel})
should determine the response in the $k$-th harmonic in the
epicyclic angle $w_1$.
In the denominator of the resonance term $m'^2-2k^2$, the 
factor $\sqrt2$ is
simply the epicyclic frequency, so that we seem to see
a resonance between the frequency with which a body in the disk 
encounters a forcing, $m\Omega$,
and a harmonic of the epicyclic frequency, $k\kappa$, which 
is a natural frequency a body wants to respond with. 
If the forcing is near resonant 
with the response times the epicyclic frequency but has a slightly
higher angular symmetry, the response is damped, whereas in the reverse
case it is amplified, a behaviour common in forced oscillations.
This resembles the situation at a Lindblad resonance, with the
difference that for one, the resonance is not exact (if it were, the
series would evaluate to infinity), and, for second,
it occurs on the entire disk.

Let us now turn to the general case of rotating modes.
As has been hinted at above, the main discomfort with
the notion of modes in the singular disk is that in its governing
equation $\omega$ factors out, so that without further devices a disk
is either stable or becomes unstable at all growth rates and pattern
speeds at once.  This seems weird enough to dismiss the possibility
of such modes.  On the other hand, this would imply that even a
completely cold disk would be stable to all non-axisymmetric
perturbations, which again would be quite
an exotic situation.

Quite recently, Goodman~\&~Evans (\cite{goodev}) have tackled 
the problem employing the Jeans equations.
The basic results of their work are that the Mestel disk's weirdness 
starts in its centre, and once one fixes boundary conditions there,
the Mestel disk becomes quite tame.  Without that boundary
condition, however, the problem is not well-posed. 
By the requirement that the
centre does not absorb or emit energy, they curb down the
two-dimensional continuum of solutions to a spectrum of one-dimensional 
continua, in
which the ratio between growth rate and pattern speed is fixed and
only the phase of $\omega$ remains unknown.  Even this ambiguity can
be removed by fixing the phase shift during the reflection of a wave
at the disk's centre, thus reducing the spectrum to a countable set of
discrete points; however, there is nothing in the equilibrium
model that would allow the definition of this phase change.

With a closed expression for the kernel of the Mestel disk, one is in a
position to reconsider the stellar-dynamical equivalent of 
Goodman~\&~Evans' problem 
using the tools of the theory of
singular integral equations:  Does the
integral equation (\ref{skernform}) admit continuous solutions, and
if, for what disk parameters?

The kernel (\ref{skernkdef}) governing the
self-consistent disk has a Cauchy-type singularity on the diagonal,
implying that the operator
$\int{\cal K}(\alpha,\alpha')\Phi(\alpha)\,d\alpha$ is not
compact, so that the familiar Fredholm theory cannot be
applied.  For the treatment of integral equations of this type,
Muskhelishvili (\cite{musk}) and his coworkers have
developed a theory. A nice summary
is found in the
more modern work of Polyanin~\&~Manzhirov (\cite{polman}).  The following
discussion is based on these works.

The integral equation (\ref{skernform}) has
the form 
\begin{equation}
0=a(\alpha')\Phi(\alpha')+{1\over i\pi}
\int_L {{\bf K}(\alpha,\alpha')\over\alpha-\alpha'}\,
\Phi(\alpha)d\alpha,\label{gencau}
\end{equation}
where ${\bf K}$ is at
least H\"older continuous, $L$ is the real axis (integration
over the real axis is henceforth implied for integral signs),
and, of course, the integral is to be understood
as a Cauchy principal value.  In the present 
case, $a(\alpha')={\cal F}(\alpha')-1$ and
${\bf K}(\alpha,\alpha')=i\pi(\alpha-\alpha'){\cal K}(\alpha,\alpha')$. The multiplication
with $(\alpha-\alpha')$ ensures that the kernel is bounded on the
diagonal, since the single zero in the denominator stemming from
$\exp\bigl(2\pi(\alpha-\alpha')\bigr)-1$ now cancels out.  The only
part of the kernel that could jeopardise the condition of H\"older
continuity is the square root, but since only $\alpha^2$ is
present
in the arguments of the square roots, ${\bf K}$ does indeed satisfy a
H\"older condition.

It is convenient to rewrite (\ref{gencau})
to the equivalent form 
\begin{eqnarray}
0&=&a(\alpha')\Phi(\alpha')
+{1\over i\pi}b(\alpha')\int{\Phi(\alpha)\over\alpha-\alpha'}d\alpha\cr
&&+{1\over i\pi}\int{\bf K}_r(\alpha,\alpha')\Phi(\alpha)d\alpha,
\label{gencaun}
\end{eqnarray}
where
\begin{eqnarray}
b(\alpha)&=&{\bf K}(\alpha',\alpha')\quad\hbox{and}\cr
{\bf K}_r(\alpha,\alpha')&=&{{\bf K}(\alpha,\alpha')-{\bf K}(\alpha',\alpha')\over\alpha-\alpha'}.
\label{cauauxfuns}
\end{eqnarray}
After this manipulation, ${\bf K}_r$ is
a compact operator, whereas 
\begin{equation}
\hat{\bf K}\Phi:=a(\alpha')\Phi(\alpha')
+{1\over i\pi}b(\alpha')\int{\Phi(\alpha)\over\alpha-\alpha'}\,d\alpha=0,
\label{charkerndef}
\end{equation}
known as the dominant or characteristic part of
the integral equation,  decides whether or not 
the integral equation has a nontrivial
solution.  For our equation, $b(\alpha')={\cal F}(\alpha')$.

To find a solution of (\ref{charkerndef}),
let us introduce an auxiliary function
\begin{equation}
\Psi(z)={1\over2\pi i}\int{\Phi_m(\alpha)\over(\alpha-z)}\,d\alpha,
\label{caupsifundef}
\end{equation}
defined for complex $z$.
If $\Phi_m(\alpha)$ is continuous -- this is henceforth assumed --,
this function is piecewise analytic in the upper
and lower half planes but may be discontinuous across the real axis.
The merit of this auxiliary function is that it allows a transformation of
the problem of the solution of the integral equation (\ref{charkerndef})
into a Riemann boundary value problem.

This can be done by applying the Sokhotski-Plemelj formulae,
that, in their formulation for the real axis, state that
when $\Phi$ and $\Psi$ are linked by (\ref{caupsifundef}),
\begin{eqnarray}
\Psi^+(\alpha')+\Psi^-(\alpha')&=&{1\over i\pi}\int{\Phi(\alpha)\over\alpha-\alpha'}\,d\alpha\cr
\Psi^+(\alpha')-\Psi^-(\alpha')&=&\Phi(\alpha')
\label{plemelj}
\end{eqnarray}
hold for all $\alpha'\in\mathbb{R}$.
Here, $\Psi^+(\alpha')$ and $\Psi^-(\alpha')$
are the limiting values of $\Psi(z)$ as $z\to\alpha'$ from above and
below the real axis, respectively.

Inserting (\ref{plemelj}) into (\ref{charkerndef}) and collecting
the terms yields 
\begin{equation}
\Psi^+(\alpha')={a(\alpha')-b(\alpha')\over
a(\alpha')+b(\alpha')} \Psi^-(\alpha')=:D(\alpha')\Psi^-(\alpha').
\label{caupreboo}
\end{equation}
Going back to the integral equation for
the singular disk, one has 
\begin{equation}
D(\alpha')=-1/(2{\cal F}(\alpha')-1).
\label{causingD}
\end{equation}
This represents a Riemann boundary value problem,
asking (in the special case relevant here) for two functions $\Psi^+$
and $\Psi^-$ that are analytic on the upper and lower half planes,
respectively, and that satisfy
$\Psi^+(\alpha')=D(\alpha')\Psi^-(\alpha')$ for all real $\alpha'$.

For reasons that will become clear below, we for the time being demand 
that $D$ has
neither zeroes nor poles on the real axis.  Comparing (\ref{somzerointeq})
and (\ref{causingD}), one sees that this will be the case exactly when
neutral modes are admitted by the disk in question, which strongly
suggests that (\ref{skernform}) has no
solution in the regular case of a bounded and nonzero $D$, and at
least one solution otherwise.
With this concept mind,
let us return to the analysis of the boundary value problem, for now keeping 
our attention on the case of a regular $D$.

The $D(\alpha')$ treated here is the restriction of an analytic
function (in particular, by assumption
it has no poles on the real axis), 
and thus is guaranteed to be H\"older continuous.
Furthermore, $D(\alpha')$ is a positive real
function for real $\alpha'$. 
This ensures that its index, the increment of its argument
over the integration contour, 
\begin{equation}
{\rm Ind}(D)={1\over 2\pi}\int d\arg
D(\alpha), \label{cauindexdef}
\end{equation}
is zero.  This in turn
implies that $\ln D$ is a well-defined (single-valued) function.
Then one can take the logarithm of both sides of (\ref{caupreboo}),
to arrive at 
\begin{equation}
\ln\Psi^+(\alpha')-\ln\Psi^-(\alpha')= \ln D(\alpha').
\label{riemannn}
\end{equation}

Using the second of the Sokhotski-Plemelj formulae
(\ref{plemelj}), it follows that with 
\begin{equation}
G(z)={1\over2\pi i}\int{\ln
D(\alpha)\over\alpha-z}\,d\alpha \label{presola}
\end{equation}
one can write
down a solution to the boundary value problem (\ref{caupreboo})
\begin{equation}
X^+(\alpha')={\rm e}^{G^+(\alpha')}\qquad X^-(\alpha')={\rm e}^{G^-(\alpha')},
\label{presolb}
\end{equation}
provided that the problem does have a solution.

Thus one has $D=X^+/X^-$, and the boundary value problem
(\ref{caupreboo}) can be written as
\begin{equation}{\Psi^+(\alpha')\over
X^+(\alpha')}= {\Psi^-(\alpha')\over X^-(\alpha')}.
\label{causolbuster}
\end{equation}
Now, the functions $G^\pm$ by construction have no
pole in their respective domains, and therefore $X^\pm$ have no zeroes.
If $\Phi^{\pm}$ are defined via an integral as in
(\ref{caupsifundef}), this implies that they are analytic in their
respective half-planes. Thus,
the functions on the two sides of (\ref{causolbuster}) are analytic
in their respective half-planes, and they are identical on the real
axis.
By Morera's theorem this implies that
the right-hand side and the left hand side represent the same analytic
function, $S(z)$, say.  Now, what are the properties of $S(z)$?

We already pointed out that $S(z)$ is analytic on the entire complex plane.
At infinity,  one has
$\Psi^{\pm}(\infty)=\pm{1\over2}\Phi(\infty)$. Since we require any
solution of the original integral equation (\ref{skernform}) to
vanish at infinity, one has $\Psi^{\pm}(\infty)=0$ as well.  Since
$D\to1$ for large $\alpha$, $G$ approaches 0 and $X^\pm$ tends to 1.
This implies that $S$ vanishes at infinity.  Since $S$ is an entire
function, it follows from Liouville's theorem 
that $S$ must be identically zero.
By the first Sokhotski-Plemelj formula one now has
$\Phi(\alpha')=S(\alpha)(X^++X^-)$, and hence the homogenous 
integral equation has no
nontrivial solution.  This is a special case of the general result
that is proven somewhat more rigorously in the literature cited above:
A homogenous Cauchy integral equation with an index of zero is
unsolvable.

Loosening the conditions on $D$, we now allow it to
have ``poles'' at some points $\alpha_i$, which, for the sake of
simplicity, are supposed to be poles of order one.  The term ``pole''
has been put in quotes since it is not necessary that $D$ is analytic
in $\mathbb{C}\setminus\{\alpha_i\}$, only that $D(\alpha_i)$ behaves
like $(\alpha'-\alpha_i)^{-1}$ at the exceptional points.

Writing
\begin{equation}
D(\alpha')={D'(\alpha')\over\prod_{i=1}^n(\alpha'-\alpha_i)},
\label{caudprime}
\end{equation}the boundary
condition (\ref{caupreboo}) takes the form
\begin{equation}
\Psi^+(\alpha')={D'(\alpha')\over\prod_{i=1}^n(\alpha'-\alpha_i)}\Psi^-(\alpha'),
\label{cauprebooprime}
\end{equation}
where $D'$ now satisfies the conditions
we required of $D$ in (\ref{caupreboo}).  Solving the boundary
value problem for this $D'$ and substituting the result back into
(\ref{cauprebooprime}), the equivalent of (\ref{causolbuster}) now
is 
\begin{equation}
{\Psi^+(\alpha')\over X^+(\alpha')}=
{\Psi^-(\alpha')\over X^-(\alpha')\prod_{i=1}^n(\alpha'-\alpha_i)},
\label{causolmaker}
\end{equation}
where $X$ and $G$ are again defined by (\ref{presolb}) and (\ref{presola}), 
except that now $D$ is replaced by $D'$.
Let us again call the analytic function defined in the left and right sides of
(\ref{causolmaker}) $S(z)$.  

By the right side of (\ref{causolmaker}), $S$
now may have poles of order 1 at $\alpha_i$ but still has to vanish
at infinity.  Applying the generalised
Liouville theorem, it follows
that any solution of the
integral equation must be of the form
\begin{equation}
\Phi(\alpha)=
\sum{c_i\over \alpha-\alpha_i}\left(
X^+(\alpha)+X^-(\alpha)\prod(\alpha-\alpha_i)\right).
\label{letdown}
\end{equation}
Evidently, solutions of this type cannot
vanish at $\infty$, since such solutions will at best converge to 
$X^-(\infty)$ times a nonzero constant as $\alpha'\to\infty$.

Thus, provided that all singularities of $S(z)$
are poles\footnote{This assumption is hard to prove, but generally true,
which is why Pipkin
(\cite{pipkin}) calls it Gentlemen's Theorem Number~1 -- a theorem no 
gentleman would question.}, the
self-consistent disk admits no growing modes at all.  
The obvious explanation for this bizarre absence of growing modes even in 
very cold disks is that
growing modes would introduce scales into the scale-free disk.

Stepping back to the mathematics for a moment, the fact that
${\cal F}$ is purely real on the real axis 
is crucial for that result, since even a
minute complex contribution could provide for modes.  For
example, adding an $i\epsilon$ to $\cal F$ would not change the result
that the index (\ref{cauindexdef}) of the problem (\ref{caupreboo})
vanishes when ${\cal F}(\alpha')<1/2$ for all $\alpha'$, so that
there still would be no modes above the stability limit of the neutral
modes.  Below that stability limit, however, the index would be
nonzero and the integral equation would admit nontrivial solutions.

Such an imaginary contribution would necessitate that the real part
of the kernel (\ref{skernkdef}) shows a singularity on the
diagonal.  We cannot offer an explanation how this might be linked to
the scale-freeness of the disk at this point.

\section{Conclusions}

In this paper we have derived an analytic expression for the kernel
of the Mestel disk.  In the cut-out disk, it reproduces the behaviour
described by Zang and ERII both qualitatively and quantitatively surprisingly
well.  It turns out that the expressions for the kernel are quite large
in the cut-out case, at least when one uses the classic cut-out functions
suggested by Zang, so the computational effort
involved in finding modes is about as large with our formalism
as it is with the older seminumerical scheme.  On the other hand, it
seems that with carefully selected cut-out functions, the kernel could
be simple enough for further analytic work.

The kernel governing modes in the self-consistent disk
is relatively compact.  It allowed us to interpret the
stability behaviour of neutral modes in terms of global resonances
between the orbital (which coincides with the frequency of excitation in
this case) and epicyclic frequencies.  In the
longstanding question of rotating modes we could establish that
no rotating modes can exist above the stability limit of the
neutral modes.  It might seem surprising that rotating modes should know
about the stability limit of the rotating modes, considering that the
ratio of orbital and epicyclic frequency is no longer independent of the
radius for rotating modes.  However, due to the lack of a length scale in the
Mestel disk, any mode is arbitrarily close to a non-rotating one.

Even below the stability limit of the neutral modes, no rotating 
or growing modes exist in the perfectly scale-free disk under reasonable
assumptions like continuity of the coefficient function in the expansion
of the potential perturbation (\ref{potexpansion}).
This bizarre property
is probably best explained by an inability of the disk to break its
self-similarity and thus by a completely artificial feature.  
In conclusion, it seems to us that the self-consistent disk's peculiar properties make its investigation
something of a pedagogical (Goodman~\&~Evans \cite{goodev}) exercise, 
whereas the cut-out disk might yet prove to be a valuable
testbed for wave dynamics in a fairly realistic full-disk setting.

\begin{acknowledgements}
      Part of this work was supported by a grant of the 
			Land Ba\-den-W\"urttemberg.  M.~D.~wishes to thank
			Jenny Read for virtual but nonetheless useful discussions.
			An anonymous referee prompted us to clarify the summary
			of Goodman~\&~Evans (\cite{goodev}).
\end{acknowledgements}


\begin{thebibliography}{}
\bibitem[1972]{abrsteg} Abramowitz, M., Stegun, I., 1972, 
Handbook of Mathematical Functions, U.S.~Government Printing Office

\bibitem[1989]{bertin} Bertin, G., Lin, C.~C., Lowe, S.~A., Thurstans, R.~P.,
1989, ApJ 338, 78

\bibitem[1987]{bintre} Binney, J., Tremaine, S., 1987, Galactic
Dynamics, Princeton University Press

\bibitem[1991]{biviano} Biviano, A., Girardi, M., Giuricin, G., 
Mardirossian, F., Mezzetti, M., 1991, ApJ 376, 458

\bibitem[1961]{bogmip} Bogoliubov N.~N., Mitropolski Y.~A., 1961, Asymptotic Methods in 
the Theory of Non-Linear Oscillations, Gordon and Breach

\bibitem[1972]{clubro} Clutton-Brock, 1972, Ap\&SS 17, 292

\bibitem[1997]{1997MNRAS.285...49C} Collett, J.~L., Dutta, S.~N., Evans, N.~W., 1997, MNRAS 285, 49

\bibitem[2000]{diss} Demleitner, M., 2000, A new approach to 
the problem of modes in Mestel disks,
Dissertation, Universit\"at Heidelberg

\bibitem[1998a]{1998MNRAS.300...83E} Evans, N.~W., Read, J.~C.~A., 1998a, MNRAS 300, 83

\bibitem[1998b]{1998MNRAS.300..106E} Evans, N.~W., Read, J.~C.~A., 1998b, MNRAS 300, 106

\bibitem[2001]{fuchspre} Fuchs, B., 2001, Density Waves in the Shearing Sheet, II: Global Modes, A\&A in preparation

\bibitem[1965]{1965MNRAS.130..125G} Goldreich, P., Lynden-Bell, D., 1965, MNRAS 130, 125

\bibitem[1999]{goodev} Goodman, J., Evans, N.~W., 1999, MNRAS, 309, 599

\bibitem[1966]{1966ApJ...146..810J} Julian, W.~H., Toomre, A., 1966, ApJ 146, 810

\bibitem[1971]{kala} Kalnajs, A.~J., 1971, ApJ 166, 275

\bibitem[1977]{kalb} Kalnajs, A.~J., 1977, ApJ 212, 637

\bibitem[1946]{landau} Landau, L.~D., 1946, J.~Phys.~USSR 10,
25 (reprinted in: Ter Haar, D., 1969, Men of
Physics: L.D.~Landau II, Pergamon Press)

\bibitem[1967]{antispir} Lynden-Bell, D., Ostriker, J.~P. 1967,
MNRAS 136, 293

\bibitem[1976]{mark} Mark, J.~W.-K., 1976, ApJ 205, 363

\bibitem[1953]{musk} Muskhelishvili, N.~I., 1953, Singular integral equations,
Noordhoff (based on the second Russian edition published in 1946)

\bibitem[1991]{persal} Persic, M., Salucci, P., 1991, ApJ 368, 60

\bibitem[1991]{pipkin} Pipkin, A.~C., 1991, A course on integral equations, 
Springer

\bibitem[1998]{polman} Polyanin, A.~D., Manzhirov, A.~V., 1998, Handbook of integral equations, CRC Press

\bibitem[1997]{1997PhDT.........1R} Read, J.~C.~A., 1997, The Stability of Model Disk Galaxies,
Ph.D.~Thesis, University of Oxford 

\bibitem[1964]{toomreq} Toomre, A., 1964, ApJ 139, 1218

\bibitem[1977]{1977ARA+A..15..437T} Toomre, A, 1977, ARA\&A 15, 437

\bibitem[1981]{toomreswing} Toomre, A, 1981, In: The structure
and evolution of normal galaxies, Cambridge University
Press, 111

\bibitem[1976]{1976PhDT........26Z} Zang, T.~A., 1976, The Stability of a Model Galaxy, Ph.D.~Thesis, 
Massachusetts Institute of Technology, Cambridge, MA

\end{thebibliography}
\end{document}